\documentclass[preprint, review,3p,times, 11pt]{elsarticle}
\usepackage{amssymb}
\journal{arXiv}
\usepackage{longtable}
\usepackage{lscape}
\usepackage{rotating}
\usepackage{multirow}
\usepackage{romannum}
\usepackage{adjustbox}
\usepackage{graphicx}
\usepackage{pgfplots}
\usepackage{comment}
\usepackage[utf8]{inputenc}
\usepackage[T1]{fontenc}
\usepackage{mathtools}
\usepackage{pdfpages}
\usepackage{pgfplots}
\pgfplotsset{compat=1.3} 
\usepackage{float}
\graphicspath{ {./img/}}
\usepackage{url}
\usepackage{amssymb}
\usepackage{pifont}
\usepackage{amsmath}
\usepackage{appendix}
\usepackage{footnote}
\usepackage{threeparttable}
\begin{document}

\begin{frontmatter}
\title{A Survey of COVID-19 Misinformation: Datasets, Detection Techniques and Open Issues}

\author[cuet]{A.R. Sana Ullah}
\ead{u1604050@student.cuet.ac.bd}
\author[cuet]{Anupam Das}
\ead{u1604054@student.cuet.ac.bd}
\author[stfx]{Anik Das}
\ead{x2021gmg@stfx.ca}
\author[csu]{Muhammad Ashad Kabir\corref{corau}}
\cortext[corau]{Corresponding author}
\ead{akabir@csu.edu.au}
\author[iit]{Kai Shu}
\ead{kshu@iit.edu}

\address[cuet]{Department of Computer Science and Engineering, Chittagong University of Engineering and Technology, Chottogram, Bangladesh}
\address[stfx]{Department of Computer Science, St. Francis Xavier University, NS, Canada}
\address[csu]{School of Computing and Mathematics, Charles Sturt University, NSW 2795, Australia}
\address[iit]{Department of Computer Science,  Illinois Institute}

\begin{abstract}
Misinformation during pandemic situations like COVID-19 is growing rapidly on social media and other platforms. This expeditious growth of misinformation creates adverse effect to the people living in the society. Researchers are trying their best to mitigate this problem using different approaches based on Machine Learning (ML), Deep Learning (DL) and Natural Language Processing (NLP). This survey aims to study different approaches of misinformation detection on COVID-19 in recent literature to help the researchers in this domain. More specifically, we review the different methods used for COVID-19 misinformation detection in their research with an overview of data pre-processing and feature extraction methods to get better understanding of their work. We also summarize the existing datasets which can be used for further research. Finally, we discuss the limitations of the existing methods and highlight some potential future research directions along this dimension to combat against the spreading of the misinformation during a pandemic.

\end{abstract}


\end{frontmatter}



\section{Introduction}\label{intro}
Coronavirus Disease 2019 (COVID-19) is an infectious disease that is caused by a newly discovered virus SARS-Coronavirus-2 (SARS-CoV-2), which is closely related to the SARS virus~\cite{Moriguchi2020}. The disease was first identified in Wuhan, China, and the first case of COVID-19 was reported on December 31, 2019~\cite{whocase}. In the beginning, it was recognized to be a global health issue, later declared as a pandemic by the World Health Organization (WHO). By October 23, 2021, the total case count crossed the bar of 243.8 Million, and over 4.9 Million people have already lost their lives worldwide~\cite{wmeter}. Owing to the brutal behavior of the virus, the sharp increase in infection and mortality led to a massive impact on several sectors such as the country's economy, public and private sectors, government bodies, and above all affecting the mental and physical health of the people by tempering their everyday lives.

During this devastating pandemic situation, people worldwide are going through an unprecedented set of challenges and fear. Every now and then, they seek information about COVID-19 solutions, e.g., medicines, vaccines, mask usage, or regarding COVID-19 dangers on various online platforms using different languages. Along with factual information, it is observed that a large amount of misinformation related to COVID-19 is circulating through these platforms. Consequently, the 'infodemic' of rumors and misinformation related to the virus came to the surface. The term 'infodemic' was first coined by the World Health Organization (WHO) meaning to an overabundance of both inaccurate and accurate information to explain the misinformation about the virus and makes it harder for people to find trustworthy and reliable sources for any claim made on any online platforms during the pandemic~\cite{WHO2020, Zarocostas2020}.

Misinformation is a piece of false information or inaccurate information that is intentionally created to get more attention from people. There are numerous terms related to misinformation including fake news, misleading news, rumors, and disinformation, which usually contain information that misguides people. During this COVID-19 situation, there has been an expeditious growth in usage of social media platforms and blogging websites which has passed 3.8 billion marks of active users~\cite{Huang2020}. People are now getting more involved in these platforms, especially on Facebook, Twitter, Instagram, etc., and expressing their thoughts, news, opinions, and information related to COVID-19. They gather information about COVID-19 from any news media or social media platforms and share it with others without fact-checking the information. As a result, it is causing panic to the people within these platforms and affecting people’s mental health, daily lives, and behaviors. 

As people's activity on social media and other online platforms has been increased significantly in this pandemic situation, the misinformation provided on these platforms easily mislead the people. So, it is now a global concern to mitigate the spread of misinformation related to COVID-19 on these platforms. It has already gained a great deal of attention from researchers all around the world. Therefore, a significant number of research works have already been done. So, it is necessary to investigate the existing studies, their findings, and also their significance. Here, we present a systematic review of various misinformation detection approaches related to COVID-19 and discuss the promising research directions.

In particular, we have made the following contributions in this survey paper.
\begin{itemize}
\item We have conducted a systematic review of existing studies on COVID-19 misinformation detection using ML techniques.
\item We have done qualitative and quantitative analysis of the selected papers by considering the datasets, pre-processing techniques, feature extraction and classification methods.  
\item Finally, we discuss the open issues and the future research directions that can help the researchers who are willing to work in this domain.
\end{itemize}


The rest of the paper is organized as follows. Section \ref{sec:overview}
provides an overview of different traditional ML and DL methods. Section \ref{sec:methodology} presents our methodology to search databases along with the selection criteria of the articles. Section \ref{sec:Analysis} outlines different datasets for COVID-19 misinformation, presents an analysis of various pre-processing, feature extraction and classification methods used in the state-of-the-art research. Section \ref{sec:open issues} discusses open issues and future research directions. Finally, Section~\ref{sec:conclusion} concludes the paper.

\section{Overview of Traditional ML and DL methods}\label{sec:overview}
ML is a subset of artificial intelligence where the main aim is to train machines by using algorithms about some statistical phenomenon to make decisions like human. It identifies the pattern of the datapoint based on some mathematical relation and predicts the new datapoint in similar way. In this section, we discuss the theoretical concepts of the ML methods which are related to our study. All the ML methods are discussed in two separate subsections named Traditional ML methods and DL methods in the following.

\subsection{Traditional ML Methods}
Traditional means the things that we have been doing for years. These traditional ML methods work as the base for the cutting edge ML methods. These algorithms learn from the data where the input features to be fed into the chosen algorithm are made by the subject matter experts. These models expects all inputs to in the format of structured data like numbers.
Traditional ML models can be used to solve classification~\cite{tradMLForClassification}, regression~\cite{genkin2007large}, clustering~\cite{finley2005supervised} etc. Here we describe the traditional ML methods that we have explored in our study.

\textbf{Logistic regression (LR)}~\cite{Cessie1992RidgeEI} is a statistical model based on the sigmoid or logistic function. It's a probability-based predictive analytic algorithm. It is a S-shaped curve which maps output value between 0 and 1 taking any real valued number as an input. The logistic regression hypothesis function is defined in~\cite{Tom} where $\beta{0}$ is the bias or intercept term and $\beta{1}$ is the coefficient for the single input value (X). This can be written as Equation~\ref{equ:LR}. 

\begin{equation}
\label{equ:LR}
    h_{\theta} (X)=\frac{1}{1 + e^{-\beta_{0}+\beta_{1}X}}
\end{equation}

\textbf{Support vector machine (SVM)}~\cite{Cortes1995} is one of the most popular and widely used algorithms for classification problems in a lot of research areas~\cite{Cristianini2008}. It creates a hyperplane or set of hyperplane in a high dimensional space for the classification of the data point based on the feature set~\cite{Cortes1995}. The dimension of the hyperplane depends upon the number of features. When the number of input features is two, the hyperplane is only a single line whereas the hyperplane becomes a two-dimensional plane if the number of input features is three.

The main objective is to identify the optimal hyperplane that differentiate the data points with maximum margin (the maximum distance between the data points of both classes) as there could be many possibilities for a hyperplane to exist in an N-dimensional space. The hyperplanes are decision boundaries that help to classify the data points. The cost function for the SVM model~\cite{Kecman2005} is written as below Equation~\ref{equ:svm}: 
\begin{equation}
\label{equ:svm}
    J{ (\theta)}= \frac{1}{2}\sum_{i=1}^{n}\theta\textsubscript{j}\textsuperscript{2}
\end{equation}

\textbf{Naive bayes (NB)}~\cite{GeorgePat} is a simple  probabilistic model based on Bayes theorem with strong independence assumption. It is the simplest form of Bayesian Network which involves a conditional independence assumption. Bayes theorem can be formulated as Equation~\ref{equ:NB}.
\begin{equation}
\label{equ:NB}
    P (\frac{A}{B})=\frac{P (\frac{B}{A})P (A))}{P(B)}
\end{equation}
 Where $A$ and $B$ are events.
 $P(A)$ and $P(B)$ are the probability of the events.
 
The \textbf{k-nearest neighbors (kNN)}~\cite{Aha1991} is one of the simplest algorithms where a datapoint is classified based on the nearest datapoints. It is a non-parametric method used for classification and regression~\cite{doi:10.1080/00031305.1992.10475879}. In this method, the euclidean distance between for each test datum and all the training data are calculated and the test datum is classified in such class that most of the k-nearest train data have and the value of \textit{k} estimates the majority of its neighbors' votes. Suppose, if $k=1$, the new datapoint is assigned to a class based on the closest training datum's class. If $k>1$, then the datapoint is assigned to a class which is the most of the neighbours' class. The mathematical formulae to estimate the Euclidean distance between two points can be calculated according to Equation~\ref{equ:kNN}:
\begin{equation}
\label{equ:kNN}
    Euclidean  Distance=\sqrt{\sum_{i=1}^{k} (x^{i}-y^{i})^2}
\end{equation}

\textbf{Decision tree(DT)}~\cite{Quinlan} is a supervised learning algorithm. It can be used for both classification and regression problems. A tree is composed of nodes, each of which represents a data feature and the relation between them represents the decision rule. The leaf of the tree represents an outcome and the values can be categorical or continuous. The choice of an attribute selection measure and a pruning method are required for designing a decision tree. There are numerous strategies for selecting attributes and the majority of them directly assign a quality measure to the attribute. Information Gain Ratio criterion~\cite{Salzberg1994} and Gini index~\cite{Breiman1983} are the most frequently used attribute selection measures in decision tree.

\textbf{Random forest (RF)}~\cite{Breiman2001} is an ensemble learning method that uses a combination of tree classifiers to perform classification, regression and other tasks. A random vector that is sampled independently from the input vector is generated for each classifier and each tree casts a vote for the most common class. This helps in the classification of an input vector~\cite{Breimanrandomforest}. Therefore, a combination of features or randomly selected features at each node make a tree. Bagging method is used to generate a training data set by randomly drawing with replacement $N$ examples, where $N$ is the size of the original training set~\cite{Breiman1996}, was used for each feature/feature combination selected. As an attribute selection measure, the random forest classifier uses the Gini index and this helps to measures the impurity of an attribute with respect to the classes.

\textbf{eXtreme gradient boosting (XGBoost)}~\cite{chen2015xgboost} is an implementation of optimized gradient boosted decision tree. It has the relatively faster computation capability in all the computing environments. It dominates tabular and structured datasets on classification and regression predictive modeling problems.This algorithm is widely used for its performance in modeling newer attributes and classification of labels. The XGBoost algorithm's evolution started with an approach focused on the decision tree where a decision is computed based on certain conditions. Sometimes single model cannot be helpful to get the correct output. For this, a systematic solution from ensemble learning can be helpful to combine the predictive power of multiple learners and it gives the aggregated output from several models. 

Bagging is one of the widely used ensemble learners. It randomly selects features and construct a forest or aggregation of decision trees. Furthermore, the model efficiency has been enhanced by reducing errors from building sequential model and the gradient decent algorithm was employed to reduce the errors in the sequential model for more improvement. Finally, XGBoost algorithm was recognized as a convenient approach by removing missing values and minimizing overfitting problems using parallel processing. It prevents overfitting problem by supporting both L1 and L2 regularization~\cite{xgboost}.  

\subsection{Deep learning methods}
Deep learning (DL) is one of the most widely explored research topics in ML which was first introduced by Rina Dechter in 1986~\cite{RinaDechter}. DL is an emerging technology that is being used in numerous applications such as Computer Vision~\cite{Diba2017,Ouyang2017}, NLP and Language Modeling~\cite{Klosowski2018}, Speech Recognition~\cite{Santana2018}, Social Network Analysis~\cite{Jin2018}, Anomaly Detection~\cite{Du2017}, Healthcare Assistance~\cite{Sharma2020}, etc.
DL has become very popular to deal with NLP tasks. Several DL methods are extensively used in this regard. Some commonly used DL methods are briefly introduced below.

\subsubsection{Simple neural network}
\begin{figure}[!htbp]
\centering
\includegraphics[width=0.7\linewidth]{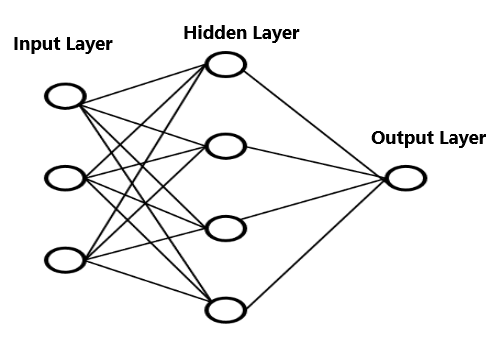}
\caption{A Simple Neural Network}
\label{NN}
\end{figure}

Neural Network (NN) refers to a system of interconnected units called neurons. A NN operates as like as the working of a human brain's neural network. A simple NN consists of three key components: input layer, hidden layer, and output layer. An input layer consists of some input neurons and represents the input values to the network. The input layer neurons are connected to the next layer neurons with some connections having particular weights. The input values get multiplied with the weights, and a bias value is added to form a value which is then passed through the activation function present in the hidden layer neurons. An activation function simply decides whether a neuron will be activated or not. A hidden layer of a neural network performs the necessary transformation of data that can be used by the output layer. Finally, there is an output layer that is responsible to generate the final result. Figure~\ref{NN} represents the architecture of a simple NN.

\subsubsection{Convolutional neural network}
Convolutional neural network (CNN) is a class of deep neural networks (DNNs) designed to process data that has a grid-like structure such as an image. It was first introduced in the 1980s for document recognition tasks~\cite{Lecun1998}. A CNN architecture consists of several layers such as an input layer, a convolutional layer with multiple filters/kernels, a pooling layer, a fully connected layer. A basic CNN architecture is illustrated in Figure~\ref{CNN}. Different layers of a CNN model are outlined as follows:

\begin{figure}[!htbp]
\centering
\includegraphics[width=0.8\linewidth]{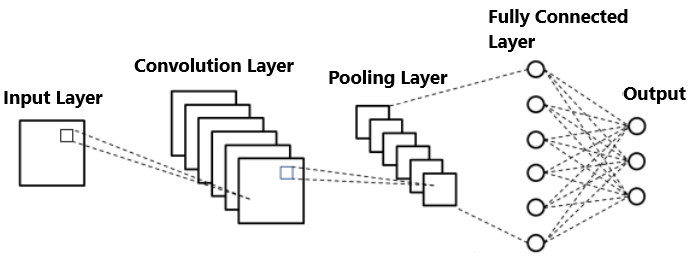}
\caption{A Basic CNN Architecture~\cite{BasicCNN2019}}
\label{CNN}
\end{figure}

i) Input layer: An input layer takes the text inputs and transforms them into a matrix form which is known as a word embedding vector. In this layer, each word of a text sequence is transformed into a dense vector of fixed size.

ii) Convolutional layer: A convolutional layer comprises several filters that perform the convolution operation on their input. A convolution is a mathematical operation that takes two inputs such as the input layer matrix and a convolution filter or kernel. A dot product is taken between the filter and the parts of the input matrix regarding the filter size by sliding the convolution filter over the input layer matrix. After the input layer matrix passes through this layer, a feature map with a single column is obtained as output. After the operation of convolution is complete, an output matrix of features is obtained through an activation function (e.g., $tanh$) upon the addition of a bias value.

iii) Pooling layer: A pooling layer performs dimensionality reduction of its input feature vectors. It uses sub-sampling to the output vectors of the convolutional layer combining neighboring elements. Different types of pooling are used in this layer such as Max Pooling, Average Pooling, etc. However, max pooling is the most common approach used in the pooling layer. In max pooling, the largest value is taken from the feature map found from the convolutional layer.

iv) Fully connected (FC) layer:
FC layers are the last layers of a CNN architecture. It can be made up of one or more layers and is placed after the pooling layers. The output of the pooling layer gets flattened before feeding it into the FC layers. Different activation functions, e.g., softmax or sigmoid are used to decide the final output of this layer. In particular, the sigmoid activation function is used in the binary classification task, whereas the softmax activation function is generally used for a multi-class classification problem.

\subsubsection{Recurrent neural network}
\begin{figure}[!ht]
\centering
\includegraphics[width=0.8\linewidth]{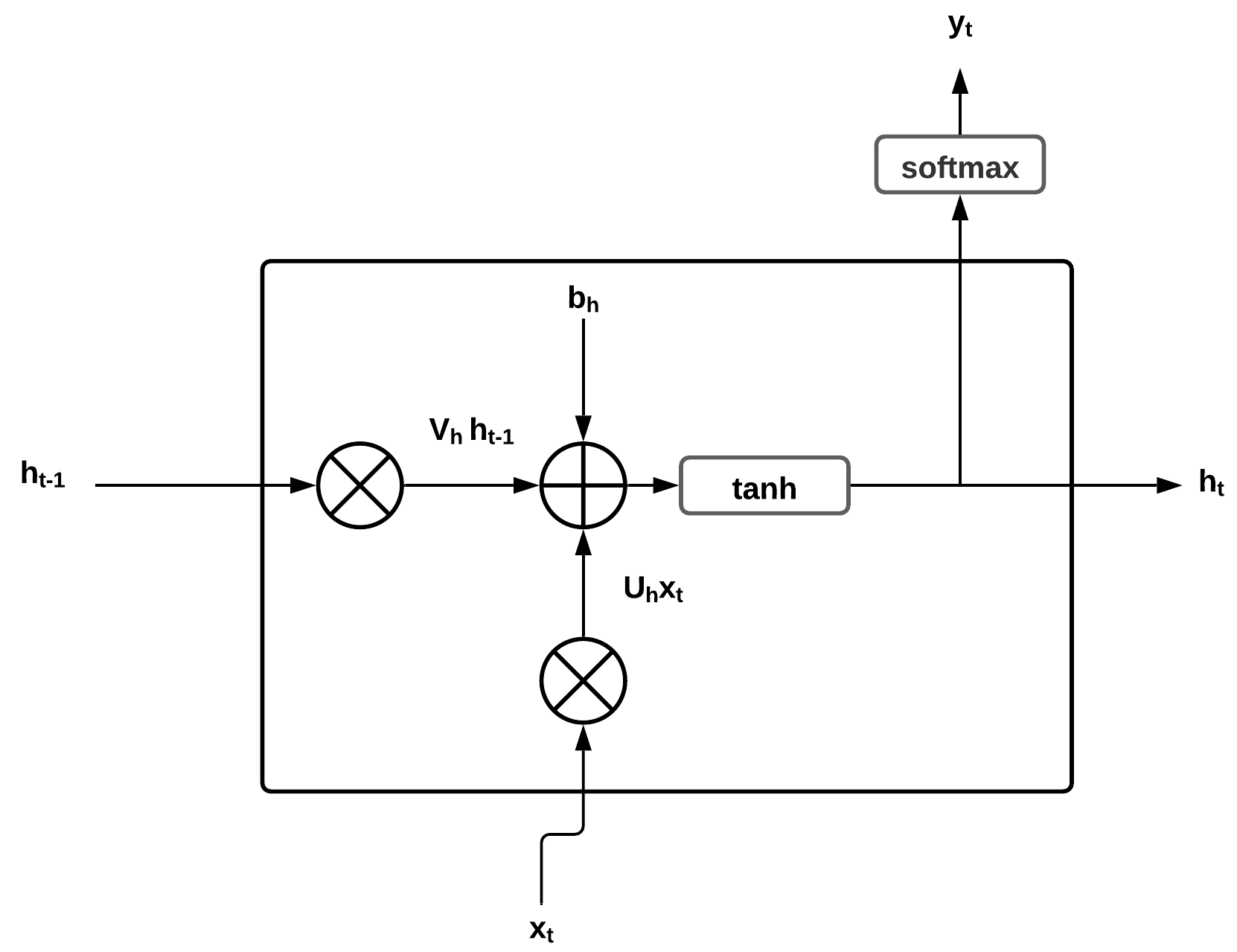}
\caption{A Basic RNN Architecture}
\label{RNN}
\end{figure}

Recurrent neural network (RNN) is a class of artificial neural networks (ANNs) that employs the sequential information in the network, which is important in the applications where the embedded structure in the data sequence conveys useful knowledge~\cite{Alkhodair2020}. In an RNN architecture, the output from the previous step is fed as input to the current step. In traditional neural networks, we presume that all inputs are independent of each other. But there are many cases where this assumption becomes quite impractical. If one wants to predict the next word of a given sentence, it is required to know the previous words of it. Consequently, RNN comes into existence. RNN performs the same operation for each word of a given sentence one after one, by taking into account its previous information. It can remember the information about a sequence over the time. While working with RNN, a sentence is considered as a sequence of words. At each timestamp, only one word is fed as input to RNN and this continues until the whole sequence is finished. After feeding the whole sequence, a corresponding output is produced at the end of the RNN model. Figure~\ref{RNN} illustrates a basic RNN architecture. At time $t$, if $x$ is given as input, we can compute the hidden state, $h\textsubscript{t}$ as of Equation~\ref{eqn:h_state}.

\begin{equation}
\label{eqn:h_state}
h_{t}  = tanh (U_{h}x_{t} + V_{h}h_{t-1} +b_{h})
\end{equation}
where $tanh$ is an activation function, $U\textsubscript{h}$ and $V\textsubscript{h}$ are the weight matrices for the current input $x\textsubscript{t}$, $h\textsubscript{t}$ and $h\textsubscript{t-1}$ stands for current and previous hidden states respectively, and $b\textsubscript{h}$ is the corresponding bias value. The output $y\textsubscript{t}$ is finally obtained by Equation~\ref{eqn:out_state}.

\begin{equation}
\label{eqn:out_state}
 y_{t}  = softmax (W_{y}h_{t}+b_{h})
\end{equation}

where $softmax$ is an activation function and  $W\textsubscript{y}$ represents the weight matrix for current input $x\textsubscript{t}$.

Problems that require learning long-term dependencies can be difficult to solve using basic RNNs. This is because a basic RNN generally suffers from the vanishing gradient problem. There are two variants of RNN such as LSTM~\cite{LSTM1997} and GRU~\cite{Chung2014} which are capable of learning long-term dependencies and can handle the vanishing gradient problem as well. LSTM uses a memory cell that can maintain information in memory for long periods of time. It involves a set of gates to control the flow of information. On the other hand, GRUs do not use separate memory cells and have fewer gates than LSTMs to control the flow of information.

\subsubsection{Bidirectional encoder representations from transformers}
\begin{figure}[!ht]
\centering
\includegraphics[width=0.8\linewidth]{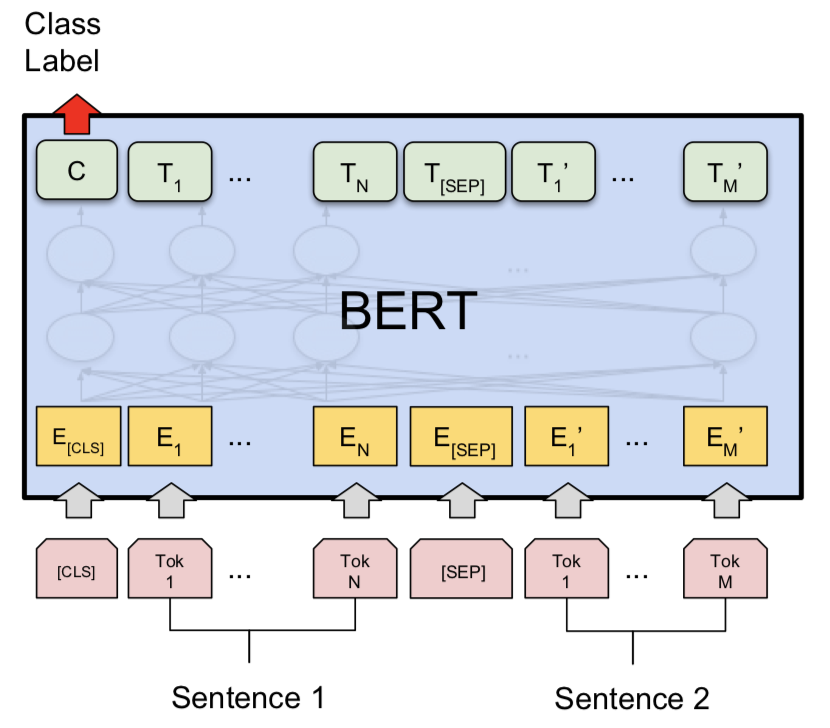}
\caption{BERT Architecture~\cite{Devlin2019}}
\label{BERT}
\end{figure}

Bidirectional encoder representations from transformers (BERT) is a transformer-based DL model that is widely used for NLP tasks. BERT is a very recent technique designed for unsupervised pre-training of deep bidirectional representations from the texts~\cite{Devlin2019}. BERT is an encoder-only transformer which can represent any token based on its bidirectional property. It can capture both left and right contexts in all layers as it is deeply bidirectional. BERT has been pre-trained on a large corpus of unlabeled data including Wikipedia and Book Corpus. Its pre-trained models are available in two sizes, e.g., BERT\textsubscript{BASE} and BERT\textsubscript{LARGE}. BERT\textsubscript{LARGE} uses higher number of parameters than BERT\textsubscript{BASE}. Figure~\ref{BERT} shows a BERT architecture for a text classification task. BERT uses an input representation by which it can represent both a single sentence and a pair of sentences in one token sequence. The first token of every input sequence is called a special classification token ([CLS]) which is useful for the classification tasks. In a single sentence task, this [CLS] token is placed in the beginning and followed by the Word Piece tokens and the separator token ([SEP]). On the other hand, for sentence pairs, the sentences are merged into a single sequence which are differentiated either by a [SEP] token or by adding a sentence embedding to each token indicating whether it corresponds to sentence 1 or sentence 2. Moreover, the corresponding positions of tokens in the sequence are represented by adding a positional embedding to each token.

BERT has been pre-trained in an unsupervised way using two tasks named Masked Language Modeling (MLM) and Next Sentence Prediction (NSP). In MLM, for a given sequence of tokens, some of them are masked. The objective is to predict only those masked tokens. In the NSP task, for a given pair of sentences, the model predicts if the second sentence logically follows the first one. BERT has inspired many variants such as XLNet~\cite{XLNet2019}, RoBERTa~\cite{Liu2019}, and ALBERT~\cite{Lan2019}. The XLNet architecture uses the bidirectional context of BERT but does not do masking. It combines the bidirectional property of BERT with auto regressive language modeling of Transformer-XL. RoBERTa is trained on more data and has longer training than BERT. It can dynamically change the masking pattern and the NSP procedure of BERT is removed here. On the other hand, ALBERT uses parameter reduction techniques to increase the training speed. To resolve the limitation of BERT regarding inter-sentence coherence, ALBERT uses sentence order prediction instead of NSP.

\section{Methodology}
\label{sec:methodology}
\subsection{Search Scope}
We have searched for articles in some prominent databases such as Scopus, Web of Science and others (e.g., Google Scholar, ResearchGate, arXiv). Scopus and Web of Science are the popular authentic databases that maintain all the published papers from IEEE, ACM, Elsevier, etc. Google Scholar, ResearchGate also provide a simple way to broadly search for scholarly literature. Furthermore, we have searched in the arXiv repository to get the preprint of the papers that are not published yet.

\subsection{Database Search Method}

We have used query string/keyword-based searching method in our study. Our query string/keyword includes COVID-19 misinformation, fake news, rumours, misleading information related studies that have used detection, classification and clustering techniques using ML techniques. The searching keywords and query strings are shown in Table~\ref{database}.
\begin{table}[!ht]
\centering
\caption{Database search string}
\label{database}
\begin{tabular}{p{2.5cm}p{12cm}}
\hline
Database name &  Query string / Keywords\\
\hline
\hline
Scopus & ( TITLE-ABS-KEY ( covid-19 )  OR  TITLE-ABS-KEY ( coronavirus ) )  AND  ( TITLE-ABS-KEY ( fake  AND news )  OR  TITLE-ABS-KEY ( misinformation )  OR  TITLE-ABS-KEY ( rumours )  OR  TITLE-ABS-KEY ( misleading ) )  AND  ( TITLE-ABS-KEY ( detection )  OR  TITLE-ABS-KEY ( classification )  OR  TITLE-ABS-KEY ( clustering ) )  AND  ( EXCLUDE ( LANGUAGE ,  "Italian" ) )  AND  ( EXCLUDE ( DOCTYPE ,  "ed" ) )  AND  ( EXCLUDE ( PUBYEAR ,  2008 )  OR  EXCLUDE ( PUBYEAR ,  1995 ) ) \\
 
\hline
Web of Science & TS=((covid-19  OR coronavirus)  AND (fake news  OR misinformation  OR misleading  OR rumours)  AND (detection  OR classification  OR clustering)) \\
 
\hline
Google Scholar, ResearchGate, arXiv  & COVID 19 fake news detection, COVID 19 fake news classification, COVID 19 misinformation detection, COVID 19 misleading news detection, COVID 19 rumour detection \\
\hline

\hline
\end{tabular}
\end{table}

\subsection{Selection Criteria Process}
For the selection of the papers for our systematic review, we have developed some eligibility criteria, which includes :
\begin{itemize}
    \item The research articles must be focused on the detection or classification of COVID-19 misinformation.
    \item The subject matter of this study exists anywhere in the title, abstract, or keywords of the article.
    \item Research papers either deploy any traditional ML and DL models to classify misinformation or represent any dataset related to COVID-19 misinformation.
    \item The articles containing classification models must have performance evaluation of the adopted methods in terms of evaluation metrics e.g., accuracy, precision, recall, F1 score, etc.
    \item The research article must be written in English.
\end{itemize}

\begin{figure}[H]
\centering
\includegraphics[width=0.9\columnwidth]{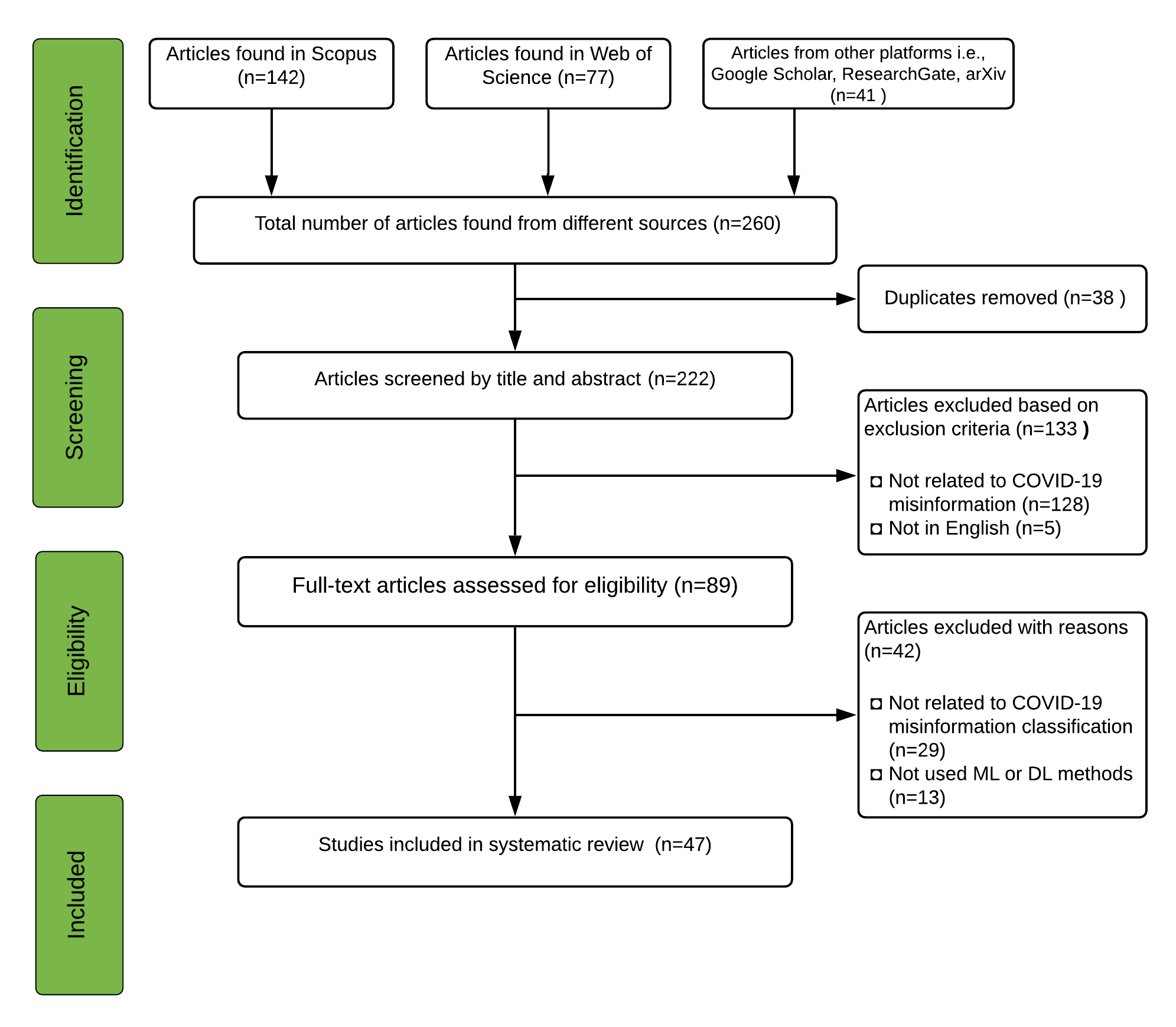}
\caption{Diagram of the systematic selection,evaluation, and quality control of the database using the Prisma model}
\label{prisma}
\end{figure}

The systematic selection process of the articles for our research is illustrated in Figure~\ref{prisma}. A total of 260 papers were found in the ``Identification phase'' of our study by searching the databases using query string/keyword. After removing 38 duplicate articles, remaining 222 articles were screened by the titles and abstracts in the ``Screening phase''. In this phase, the articles are further filtered with the eligibility criteria and 133 articles were removed. In the ``Eligibility phase'', full-texts of the remaining 89 articles were studied for final selection. 42 articles were excluded in this phase for not having relevant outcomes, results or lack of evaluation metrics. Finally, in the ``Included'' phase, we have found 47 papers which were included in our survey.

\section{Analysis}
\label{sec:Analysis}
In this section we discuss about the datasets, different pre-processing, feature extraction, and classification methods used in the existing literature regarding COVID-19 misinformation detection along with their evaluation results.

\subsection{Dataset Description}
Relevant and sufficient training data is considered as the basis to achieve precise results from any ML-based misinformation detection systems. To perform the misinformation classification task, data from various platforms such as social media, news websites, fact-checking sites, government or well-recognized authentic websites are being used frequently. But manually determining the veracity of news is a challenging task, usually requiring annotators with domain expertise who performs careful analysis of claims and additional evidence, context, and reports from authoritative sources. Therefore, to facilitate future research related to the COVID-19 misinformation task, the  datasets existing in the literature are discussed in the next.

 \subsubsection{Labeled Data}
 Datasets used in the papers~\cite{Elhadad2020,Chen2020,Elhadad2021f,Kar2020,Shahi2020,Alkhalifa2020,Koirala2020,Carlos2020,Hossain,Song2020,Boukouvalas2020,Al-Rakhami2020f,Dutta2020,Kumar2020,Patwa2020,Alsudias2020,Alam2020,Dharawat2020} are labeled in two or more classes to classify the COVID-19 misinformation using ML algorithms. The papers ~\cite{Elhadad2020,Elhadad2021f,Kar2020,Alkhalifa2020,Carlos2020,Hossain,Boukouvalas2020,Al-Rakhami2020f,Dutta2020,Kumar2020,Alsudias2020,Alam2020, Dharawat2020} use the data that are mainly collected from Twitter platform by using the Twitter APIs. On the other hand, paper~\cite{Chen2020} utilizes the data collected from various Chinese rumor refuting platforms, paper~\cite{Shahi2020} uses data that is collected from different fact-checking websites after obtaining references from Poynter and Snopes, paper~\cite{Koirala2020} makes use of the data that is scraped by Webhose.io API from various news and blog sites around the world, paper~\cite{Song2020} uses the data collected from IFCN Poynter website and paper~\cite{Patwa2020} uses the data which is gathered from public fact-verification websites and other sources e.g. World Health Organization
 (WHO), Centers for Disease Control and
 Prevention (CDC), etc. Datasets used in the articles~\cite{ Cui2020,Perrio2020,Micallef2020,Zhou2020,Memon,Haouari2020,Lee2020,Yang2020,SHAHI2021,Dimitrov2020} are labeled and publicly available which can be utilized in future research related to  COVID-19 misinformation detection and classification task. Most of the datasets contain data only in English language , except the datasets~\cite{Elhadad2020,Elhadad2021f,Haouari2020,Alsudias2020,Yang2020}. The datasets~\cite{Haouari2020,Alsudias2020} contain the tweets only in Arabic language while the dataset~\cite{Elhadad2020,Elhadad2021f} contains data in two different languages (e.g. , English or Arabic) separately. The dataset~\cite{Yang2020} contains microblogs related to COVID-19 in Chinese language. Some studies ~\cite{Kar2020,Shahi2020,Alam2020} also introduced  multilingual datasets containing data in multiple languages. Among the datasets that have already been used in existing works, the datasets~\cite{Chen2020,Koirala2020,Song2020,Al-Rakhami2020f,Dutta2020,Kumar2020} have not been made publicly available.

 \subsubsection{Unlabeled Data}
 To encourage future research on COVID-19 misinformation detection, we have collected some more datasets that are vast in size. All of these datasets ~\cite{Lamsal2020,Qazi2020,Banda,Alqurashi2020,Twitter,Lopez2020} are publicly available to use. These data are collected from the Twitter platform. The datasets ~\cite{Qazi2020,Banda,Twitter,Lopez2020} are multilingual while other two datasets~\cite{Lamsal2020,Alqurashi2020} are monolingual containing data in English and Arabic respectively. After making some modifications and proper annotations , future research works can be conducted in this domain by utilizing these datasets.
 
\subsection{Dataset pre-processing}
Data pre-processing is one of the significant parts before feeding the data in any ML algorithm. Data pre-processing includes data cleaning, normalization, transformation, feature extraction \& selection, etc. This step aims to facilitate data manipulation, reduce memory space needed, and shorten the processing of huge amounts of data. Some commonly used pre-processing techniques are described in the following.

\subsubsection{Tokenization}
Tokenization process splits the text data into smaller parts known as tokens and it removes all the punctuations from the textual data~\cite{Mullen2018}. By using tokenization process, texts can also be converted into lowercase or uppercase.

\subsubsection{Stopword Removal}
Stops Words are the most common words in a language which do not provide much context and hold less useful information. These words help to make sentence structures. Stopwords are mainly articles, prepositions and conjunctions and some pronouns for example an, are, as of, on, or, that, the, these, this, too, was, what, when, where, who, will, etc.

\subsubsection{Stemming}
Stemming is a technique that is used to convert a word to its grammatical roots so that they can be presented in one term only~\cite{inproceedings}. For example, the words ``Logically”, ``Logic” and ``Logicality” can be reduced to the word ``Logic”. Stemming is used to make classification faster and efficient. It reduces the input dimension which creates better possibility to get better accuracy.
 
\subsection{Feature Extraction}
Feature extraction is a process of selecting relevant features without losing any important information. In the text categorization, generally a document consists of a large number of words, phrases which creates a high computational burden in the learning process. Also, it is difficult to learn from high dimensional data. Besides, classifier’s accuracy can decrease for taking irrelevant features. By taking relevant and important features can help to speed up the learning process. We have found different feature extraction methods in our study which are represented in Table~\ref{embedding} and described in the following subsections.

\subsubsection{PCA}
PCA is dimensionality reduction technique. The goal of PCA is to produce lower-dimensional feature sets from the original dataset. In PCA it is very important to determine the number of principal components. If p is the number of principal components to be chosen among all of the components, the values of p should represent the data at their very best.\par

\subsubsection{ICA}
ICA is a linear transformation method in which the desired representation is the one that minimizes the statistical dependence of the components of the representation. It doesn't focus on mutual orthogonality of the components and the issue of the variance among the datapoints.
 
 \subsubsection{BoW}
The BoW model is a simplifying representation used in NLP and information retrieval (IR), where a text is represented as the bag (multiset) of its words, disregarding grammar and even word order, but keeping multiplicity. The occurrence of each word is taken as a feature.

\subsubsection{TF-IDF}
The main idea of TF-IDF comes from the theory of language modeling where the terms in a given document can be divided into two categories: those words with eliteness and those words without eliteness~\cite{article}. TF-IDF is measured by multiplying two metrics where one represents how many times a word appears in a document and the other represents the inverse document frequency of the word across a set of documents.

\subsubsection{LIWC}
LIWC~\cite{liwc} stands for Linguistic Inquiry and Word Count. It is a widely-accepted psycholinguistic lexicon. If a news article is given, LIWC can count the words in the text that fall into one or more of 93 linguistic, psychological and topical categories based on which 93 features are derived and classified within a traditional statistical learning framework.~\cite{xinyi}

\subsubsection{RST}
RST~\cite{william} stands for Rhetorical Structure Theory. It describes relations that hold between parts of text and organizes a piece of content as a tree. In the study~\cite{Zhou2020},the authors used a pretrained RST parser~\cite{ji} to obtain the tree for each news article. It counted each rhetorical relation within a tree and classified in a traditional statistical learning framework.

\subsubsection{ELMo}
ELMo stands for Embeddings from Language Models. It is a deep contextualized word representation instead of using a fixed embedding for each word developed in 2018 by AllenNLP~\cite{peters-etal}. It uses a deep, bi-directional LSTM model to create word representations. Unlike other traditional word embeddings such as word2vec and GLoVe, ELMo analyses words within the context that they are used rather than a dictionary of words or their corresponding vectors. Therefore, the same word can have different word vectors under different contexts. 

\subsubsection{Word2vec}
Word2vec is a word embedding technique developed by a team of researchers led by Tomas Mikolov at Google which uses shallow neural network~\cite{Mikolov}. There are two types of Word2Vec, Skip-gram and Continuous Bag of Words (CBOW). CBOW method takes the context of each word as the input and tries to predict the word related to the context. It has better representations for more frequent words. On the other hand, the distributed representation of the input word is used to predict the context in the Skip-gram model which works well with small amount of data and is found to represent rare words well.

\subsubsection{FastText}
FastText is an extension of word2vec model developed by Facebook AI research lab. It treats each word as n-gram of characters. So the vector for a word is made of the sum of this character n grams. It can derive word vectors for unknown words by taking morphological characteristics of words even if a word isn’t seen during training. So it works well with rare words and can provide any vector representation. 
\subsubsection{BERT}
Bidirectional Encoder Representations from Transformers (BERT) is a technique for NLP pre-training developed by Google~\cite{Devlin2019}. It is deeply bidirectional because it learns text representation for both directions for better understanding of the context and relationship. BERT developers created two main models: BERT Base and BERT Large, as well as languages such as English, Chinese, and a multi-lingual model(mBERT) covering 102 languages trained on Wikipedia.

\subsubsection{Glove}

GloVe stands for global vectors for word representation developed by Stanford as an open source project~\cite{pennington}. It is an unsupervised learning algorithm for generating word embeddings. Here, all the words are mapped into a meaningful space where the distance between words is related to semantic similarity. Training is performed on aggregated global word-word co-occurrence matrix from a corpus, and the resulting representations show interesting linear substructures of the word in vector space. 
 
\begin{longtable}[c]{p{5cm} p{6cm}}
 \caption{Feature extraction methods \label{embedding}}\\
 \hline
 \textbf{Methods}&\textbf{ Papers}\\
 \hline
 \hline
 \endfirsthead
 \caption{Feature extraction methods (continued)}\\
 \hline
 \endhead

 \hline
 \endfoot

 \hline

 \endlastfoot
 BERT& \\
 \quad Pre-trained BERT &~\cite{Kar2020,Alkhalifa2020,Hossain,Song2020,Dharawat2020}\\
 \quad mBERT&~\cite{Kar2020}\\
 \quad COVID-Twitter-BERT &~\cite{Alkhalifa2020,Hossain,Dharawat2020}\\
 GloVe& ~\cite{Cui2020,Elhadad2021f,Koirala2020,Hossain,Kumar2020,Dharawat2020}\\
 ELMo  & ~\cite{Alkhalifa2020}\\
 
 Word2Vec&~\cite{Alsudias2020}\\

FastText &~\cite{Alsudias2020,Alam2020}\\

BoW &~\cite{Cui2020,Elhadad2020,Koirala2020,Carlos2020,Hossain}\\

COUNT VECTOR &~\cite{Alsudias2020}\\

TF &~\cite{Elhadad2020}\\

TF-IDF &~\cite{Elhadad2020,Alkhalifa2020,Koirala2020,Carlos2020,Hossain,Patwa2020,Alsudias2020,Dharawat2020,Perrio2020}\\

PCA&~\cite{Boukouvalas2020}\\

ICA & ~\cite{Boukouvalas2020}\\
 
LIWC &~\cite{Carlos2020,Zhou2020}\\
 
RST &~\cite{Zhou2020}\\

 \end{longtable}

\subsection{Classification Methods}
In case of classification task, two types of classification strategies are commonly used by researchers i.e., Binary classification and Multi-class classification. As can be seen from Table~\ref{class}, binary classification is the mostly used classification strategy for classifying COVID-19 misinformation, rather than multi-class classification.

{\tiny 
\begin{table}[h!]
\begin{center}
\caption{Classification strategies followed by different methods}
\label{class}
\begin{tabular}{p{4cm} p{10cm}} 
  \hline
Classification Strategies & Papers \\ 
  \hline
  \hline
 Binary Class & ~\cite{Cui2020,Elhadad2021f,Kar2020,Shahi2020,Alkhalifa2020,Koirala2020,Carlos2020,Boukouvalas2020,Elhadad2020,Al-Rakhami2020f,Patwa2020,Dutta2020,Alam2020,Zhou2020,Perrio2020}\\
 
 Multi-Class & ~\cite{Chen2020,Hossain,Song2020,Kumar2020,Alam2020,Dharawat2020} \\ 
  \hline
\end{tabular}
\end{center}
\end{table}
}

The classification methods used in the existing works on COVID-19 misinformation detection are reviewed here in detail according to the categories described in Section~\ref{sec:overview}.

\subsubsection{Traditional ML Methods}

Traditional ML methods perform very well in the detection of misinformation based on COVID-19. Several traditional ML techniques such as kNN, Naive Bayes, Logistic Regression, SVM etc. have been used to perform this classification. Based on the ICA model, Boukouvalas et al.~\cite{Boukouvalas2020} proposed a data-driven solution where knowledge discovery and detection of misinformation are achieved jointly. Their proposed method helps to generate low dimensional representations of tweets with respect to their spatial context and deployed SVM by using different kinds of popular kernel methods, e.g, SVM/Gaussian, SVM/RBF, SVM/Polynomial. Using SVM/Gaussian model, an accuracy of 81.2\% was reported. Perrio et al.~\cite{Perrio2020} proposed a Linear SVM model using TF-IDF features due to the established high accuracy of this model on the task of text classification which achieved 81.55\% F1-score.

Elhadad et al.~\cite{Elhadad2020} proposed a voting ensemble ML classifier using ten classification algorithms (DT, MNB, BNB, LR, kNN, Perceptron, NN, SVM, RF, and XGBoost) when using TF and TF-IDF with (character
level, Unigram, Bigram, Trigram, and N-gram word size), and word embedding as feature extraction techniques.
In this study, the highest accuracy using kNN is 99.36\% which was achieved selecting character level features using 5-fold cross validation with a significant precision and recall (99.47\% and 99.73\%) respectively.

Carlos et al.~\cite{Carlos2020} presented a model for detecting COVID-19 misinformation videos on YouTube by leveraging user comments. For classifying user comments, they trained some models where two models are trained as baselines. One is LR model based on LIWC’s lexicon-derived frequencies~\cite{Tausczik} as features and other is MNB model using BOW as features. They classified the set of YouTube videos by using the percentage of conspiracy comments of each video as a feature and extracted content features from the videos’ titles and from the raw first hundred comments per video. They set six features(title, conspiracy, comments and their combination) and used LR, SVM and RF where SVM model trained using different kernel method such as linear, sigmoid, and RBF kernel. Among all the six features, comments with conspiracy features got slightly better accuracy.

In another study, the author used kNN based classifier method to find the truthfulness of the news shared on social media using their own collected dataset during four months of lockdown~\cite{Dutta2020}. Before fitting into the classifier, they preprocessed the dataset based on the similarity news in social medias. They got a decent accuracy using this classifier. In another study, this kNN classifier was used as candidate weak-learners during the experimental phase of ensemble learning where this algorithm obtained an accuracy of 94.39\% for 10 fold cross validation~\cite{Al-Rakhami2020f}.

Cui et al.~\cite{Cui2020} performed different kinds of simple methods on their own created dataset as baselines for the comparative analysis of misinformation detection task. They used BOW features and fed the representations to a linear kernel SVM and RF classifier. For feeding into the LR model, they concatenate all the word embeddings together. Although  these models didn't achieve a good score in this dataset, the comparative analysis helped to find the overall model performance. In another study~\cite{Zhou2020}, extensive experiments are conducted using ReCOVery dataset which included the baseline performances using either single-modal or multi- modal information of news articles for predicting news credibility and allow future methods to be compared to. Different kinds of methods such as : LR, NB, kNN, RF, DT  and SVM are adopted in their experiment using LIWC and RST feature. Dharawat et al.\cite{Dharawat2020} performed experiments with several  multiclass classification models on their own created new benchmark dataset- "Covid-HeRA". They use RF, SVM, LR model with BOW and  100-dimensional pretrained Glove embeddings and achived a very good accuracy above 95\%. In the study~\cite{Patwa2020}, the author performed experiment  with their annotated benchmark dataset using four ML baselines - DT, LR, GB, and SVM and obtained the best performance of 93.46\% F1-score with SVM using TF-IDF feature.

\subsubsection{DL Methods}
Over the last few years, DL is playing a vital role in misinformation detection tasks. Various DL techniques have been used to conduct the classification task of misinformation in the pre-COVID situation. During this COVID-19 situation, DL has emerged as one of the significant technologies to make efficient systems that can detect and classify the misinformation related to COVID-19. Several DL methods have been adopted in the existing research works of COVID-19 misinformation detection and classification task. These methods are reviewed here in detail according to the categories described in Section~\ref{sec:overview}.

\noindent\textbf{Simple NN: }
Neural Networks (NNs) are the most basic architectures among the DL methods. Some studies on COVID-19 misinformation detection employed NN and its previous versions in their studies. For instance, Elhadad et al. implemented the Perceptron and NN model with different feature extraction techniques such as  TF, TF-IDF, Word Embedding to construct a voting ensemble system  ~\cite{Elhadad2020}. In this study, the authors proposed an ensemble system that takes the results from Perceptron and NN model and finally performs hard voting on their results to classify the misleading information on COVID-19. They achieved high-performance scores from these two classification algorithms. The Perceptron model showed an accuracy of 99.57\% with an F1 score of 99.73\%. On the other hand, NN achieved 99.68\% accuracy with an F1 score of 99.80\%. In another study, the authors deployed Multi-layer Perceptron (MLP) model with pre-trained BERT embedding and the NN model with multilingual BERT (mBERT) embedding for the classification of COVID-19 fake tweets in multiple Indic-Languages (e.g., Hindi, Bengali) along with English language~\cite{Kar2020}. The MLP model didn't show good performance due to the smaller size of data. But, the NN model was able to deal with the small data size problem and achieved more than 80\% F1 scores in both monolingual (for English ) and multilingual (for English, Hindi, and Bengali) settings.

\noindent\textbf{CNN: }
CNN is one of the most popular and widely used models in NLP tasks. Similarly, some of the existing studies  on COVID-19 misinformation classification also adopted CNN and it's other variants for the classification purposes. For example, Cui and Lee implemented a CNN model for detecting COVID-19 healthcare misinformation~\cite{Cui2020}. They used word embedding initialized by Glove and fed it into the CNN model. In another study, the authors deployed a CNN model using  pre-trained Glove embedding to build up a system for detecting misleading information related to COVID-19~\cite{Elhadad2021f}. They utilized the word-level representation of features to preserve their order and were able to obtain high accuracy in results. Alkhalifa et al. introduced a CNN-based classification system with different pre-processing approaches and embedding methods to classify the COVID-19 rumors~\cite{Alkhalifa2020}. In this work, the best performing model comprises a CNN model with COVID-Twitter-BERT (CT-BERT) embedding which is pre-trained on COVID-19 Twitter data. Another study applied a CNN model with an embedding layer in front of it for the classification of fake news related to COVID-19 ~\cite{Koirala2020}. This study reported that lower weights of minority class cause overfitting problems. By increasing the weights of the minority class, the author was able to reduce the overfitting problem significantly and increased the test accuracy as well. Dharawat et al. introduced a dataset for health risk assessment of COVID-19 misinformation~\cite{Dharawat2020}. The authors also experimented with CNN to classify the misinformation categories using both binary and multi-class classification methods. They implemented CNN with multiple kernels and used pre-trained Glove embedding as an initialization of word embedding. Among all the studies that experimented with CNN, the study~\cite{Elhadad2021f} achieved the highest performance with CNN by reporting the accuracy and F1 score of 99.999 \% and 99.966 \% respectively. In some  studies, authors used TextCNN, a CNN architecture for text classification, to classify COVID-19 rumors~\cite{Chen2020}, fake news~\cite{Zhou2020} and  misinformation~\cite{Kumar2020} in COVID-19 tweets.The TextCNN model uses a one-dimensional convolution layer and max-over-time pooling layer to capture the associations between the neighboring words in texts. The study~\cite{Chen2020} obtained the highest performance with accuracy and F1 score of 98.40\% and 97.24\% respectively, among the studies that adopted the TextCNN model.

\noindent\textbf{RNN: }
RNN has the ability to capture better contextual information from the texts, therefore various studies utilized RNN and it's other variants for the classification of COVID-19 misinformation. In particular, Chen used the TextRNN model to classify COVID-19 rumors~\cite{Chen2020}. The author used LSTM layers inside to implement this model. In this study, higher accuracy was obtained in the classification results as TextRNN was able to capture the relationship between the semantics and the contexts strongly. As LSTM has the advantage of learning long term dependencies over RNNs, some studies implemented the LSTM  model for the better classification of misinformation related to COVID-19 ~\cite{Koirala2020, Boukouvalas2020, Kumar2020}. The study~\cite{Koirala2020} reported the best performance score using LSTM with an accuracy of 75\%. Some of the studies applied the BiLSTM model which is an extension of the LSTM architecture. A BiLSTM model can learn long term dependencies and reserve contextual information both in the forward and backward directions. Hossain et al. used the BiLSTM model to classify tweet-misconception pairs related to COVID-19~\cite{Hossain}. Other studies implemented BiLSTM for the classification of COVID-19 misinformation~\cite{Boukouvalas2020,Kumar2020, Dharawat2020}. BiLSTM using pre-trained Glove embedding came up with an accuracy of 96.6\% which is the highest among the studies that employed LSTM for the classification purpose~\cite{Dharawat2020}. Moreover, one study deployed a model called BiGRU for the classification of healthcare misinformation related to COVID-19~\cite{Cui2020}. BiGRU is a variant of RNN that consists of two GRU models. Like BiLSTM, it can also learn long-term dependencies in both forward and backward directions with only the input and forget gates. The authors used word embeddings to the BiGRU model which was initialized by Glove embedding. But they didn't achieve good results using BiGRU due to their imbalanced data.

\noindent\textbf{BERT: }
BERT is a newer DL method that has been extensively used for dealing with NLP tasks. Several exiting studies focused on BERT and its variants for classification purposes. For instance, Chen proposed a fine-grained classification method based on the BERT pre-training model to classify the rumors of COVID-19~\cite{Chen2020}. The author fine-tuned the pre-trained BERT model for classification purpose. This study demonstrates that the multiheaded attention mechanism used in BERT is capable to produce outstanding results. This study reported an accuracy of 99.20\% in the classification results using the BERT model. Alam et al. proposed a multilingual model called mBERT to analyze the COVID-19 disinformation~\cite{Alam2020}. The authors trained this model with combined English and Arabic tweets. They achieved good performance scores in both monolingual and multilingual settings using mBERT model. Shahi and Nandini performed a BERT-based classification of real or fake news on COVID-19 by introducing a multilingual cross-domain dataset~\cite{Shahi2020}. Kumar et al. introduced a dataset for fine-grained classification of misinformation in COVID-19 tweets~\cite{Kumar2020}. The authors also applied several transformer language models including three variants of BERT model (e.g., Distil-Bert, BERT-base, and BERT-large), three variants of RoBERTa model (e.g., Distil-RoBERTa, RoBERTa-base, RoBERTa-large), and two variants of ALBERT model  (e.g., Albert-base-V2, Albert-large-V2) to perform a systematic analysis. They performed fine-tuning on these pre-trained models to get them ready for their classification task. Among all the adopted models, Roberta-large appeared the best performing model with an F1 score of 76\% as it was trained on a larger corpus compared with the other models. In another study, the authors fine-tuned three transformer models e.g., XLNet base, BERT base, and RoBERTa base for the classification of user comments associated with COVID-19 misinformation videos~\cite{Carlos2020}. Among these models, RoBERTa showed the best performance in test data. 

Hossain et al. employed Sentence-BERT(SBERT)~\cite{Reimers2020} and SBERT(DA) models for their classification purpose~\cite{Hossain}. The SBERT model is a modification of pre-trained BERT architecture that uses siamese and triplet networks to extract semantically meaningful sentence embeddings. On the other hand, SBERT(DA) uses the SBERT representation with domain adaptive pretraining on COVID-19 tweets. In this work, the authors utilized COVID-Twitter-BERT embedding for domain adaptation purpose. A study showed the classification of  COVID-19 disinformation both in English and Arabic languages adopting binary and multiclass classification settings~\cite{Alam2020}. The authors finetuned the pre-trained BERT, RoBERTa, and ALBERT model for English language experiments. BERT outperformed all other models in case of English language. For Arabic language experiments, they employed AraBERT ~\cite{AraBERT2020} model which is pre-trained on a large corpus of 70 million Arabic sentences. Due to the smaller size of Arabic dataset used in the training, AraBERT didn't perform very well in their study. Besides, Perrio and Madabushi fine-tuned the  BERT\textsubscript{BASE} and RoBERTa\textsubscript{BASE} to be used as baseline models~\cite{Perrio2020}. The authors trained the BERT model with raw training data and obtained a good performance against the validation data. But the RoBERTa model achieved higher performance than BERT in the baseline results. Due to achieving higher performance using RoBERTa, the authors proposed some ensemble models to increase the performance of the baseline models using RoBERTa model. The RoBERTa+PROB+TFIDF model comprising RoBERTa with both TF-IDF features and a percentage metric of probability achieved the highest performance among all the proposed ensemble models. A F1 score of 91.51\% was reported using this model. Some other studies experimented with BERT for the classification of COVID-19 fake news ~\cite{Koirala2020}, COVID-19 disinformation ~\cite{Song2020} and COVID-19 misinformation~\cite{Boukouvalas2020,Dharawat2020}.

\noindent\textbf{Other Methods: }
Apart from the above methods, researchers also applied other DL methods for the classification purposes. For example, Song et al. adopted two different methods namely, 
SCHOLAR~\cite{Scholar2018} and NVDM~\cite{Miao2016} for topic modeling task based on the classification of COVID-19 disinformation~\cite{Song2020}. These models use the functionality of the VAE framework in document modeling tasks. VAE can make strong assumptions concerning the distribution of latent variables in the case of latent representation learning. In this work, the authors used two versions of the NVDM model e.g., NVDMo and NVDMb where NVDMo represents the original NVDM architecture and  NVDMb represents the NVDM with BERT representation. In topic modeling, SCHOLAR achieved a higher coherence score than NVDM but in terms of perplexity, NVDM showed higher performance than SCHOLAR.


Some studies employed attention-based models for the classification of COVID-19 misinformation~\cite{Cui2020, Dharawat2020}. The authors used two models based on attention mechanism namely HAN~\cite{HAN2016} and dEFEND~\cite{dEFEND2019} for their purposes. HAN uses two levels of attention mechanisms applied at the word and sentence level to learn the hierarchical structure of the documents.  It uses a bidirectional GRU network for word and sentence level encoding procedures. An attention mechanism is used after the word encoder to extract the contextually important words and form a sentence vector by aggregating the representations of the informative words. A sentence encoder then works on the derived sentence vectors and generates a document vector. Another attention mechanism is used after the sentence encoder to measure the importance of sentences in the classification of a document. The dEFEND framework utilizes the  HAN on article content and a co-attention mechanism between article content and user comments to classify misinformation. In the studies~\cite{Cui2020, Dharawat2020}, dEFEND showed higher performance scores than HAN due to its robustness and explainability as well. 

In one study, multi-modal information ( e.g., textual and visual ) of new articles on coronavirus was used for the detection of fake news ~\cite{Zhou2020}. The authors adopted the SAFE~\cite{SAFE2020} model which can jointly learn the textual and visual information along with their relationships to detect fake news. In SAFE architecture, a Text-CNN model is used to extract the textual features from the news articles and the visual features (e.g., images) are also extracted by the Text-CNN model while the visual information within the articles is first processed using a pre-trained image2sentence model. The authors achieved the best performance using the SAFE model among all the baseline methods employed.

Another study employed a model called SAME ~\cite{SAME2019} for the classification of healthcare misinformation on COVID-19~\cite{Cui2020}. SAME is a multi-modal system which uses news image, content, user profile information as well as users' sentiments to detect fake news. In this study, the authors skipped the visual part of the SAME model for their classification purpose as the majority of the news articles doesn't contain any cover images. They weren't able to get satisfactory results with this model as their dataset was quite imbalanced. 

Some other methods such as XLM-r, FastText were used to perform fine-grained disinformation analysis on Arabic tweets ~\cite{Alam2020}. In this study, the authors used these two models in both binary and multi-class classification settings. They achieved consistent and good results using FastText while XLM-r didn't perform well as the amount of data was small and it was likely to overfit.

\subsubsection{Combined Methods}
Some research works also used different combinations of traditional ML and DL methods to increase the overall performance of classification methods.

\textbf{ML+ML :} A study proposed an ensemble-learning-based framework for justifying  the credibility of a vast number of tweets based on tweet-level and user-level features~\cite{Al-Rakhami2020f}. For this, they integrated  six traditional ML algorithms utilizing stacking-based ensemble learning which resulted in higher accuracy and a more generalized model. They carried out various experiments when constructing the ensemble model. For a level-0 weak-learner, they used the SVM+RF models, while for a level-1 weak learner, they used the C4.5 model as a meta-model. Other ensemble models are : C4.5+RF, C4.5+kNN, SVM+kNN, SVM+BayesNet+kNN and C4.5+Bayes Net+kNN.

\textbf{DL+DL :} In a study, a hybrid DL model called CSI ~\cite{CSI2017}  was adopted to detect COVID-19 misinformation ~\cite{Cui2020}. CSI explores news content, user responses to the news, and the sources that users promote for the detection of fake news. In this work, the authors' utilized word embeddings initialized by the Glove embedding. Due to the imbalanced data, CSI couldn't achieve satisfactory results in this study. Elhadad et al. used a RCNN model which combines the properties of RNN and CNN to detect COVID-19 misleading information ~\cite{Elhadad2021f}. In the RCNN architecture, a recurrent structure captures the contextual information and the max-pooling layer can automatically judge which words play key roles to capture the key components in texts~\cite{RCNN2015}. In this study, RCNN performed very well and an accuarcy of 99.997\% was reported. Kumar et al. proposed a CNN-RNN model (CNN layer stacked over the RNN layer) and a RNN-CNN model (a single BiLSTM layer is employed over the top of a 1D-CNN layer) with an word embedding in the first layer for the classification of misinformation in COVID-19 tweets~\cite{Kumar2020}. Another study represents three DL models (e.g., RNN-LSTM, RNN-GRU, BiRNN-GRU) which are the combinations of various recurrent neural networks~\cite{Elhadad2021f}. These models use pre-trained Glove embedding in the first layer of each model and together constitute an ensemble DL system for detecting COVID-19 misleading information. The authors achieved very high performance from these models with more than 99\% accuracy in every cases.

Hossain et al. proposed a system that uses combinations of BERTSCORE (DA) with BiLSTM and BERTSCORE (DA) with SBERT (DA) models for detecting COVID-19 misinformation on social media~\cite{Hossain}. BERTSCORE (DA) represents BERTSCORE~\cite{Zhang2019} with domain-adaptive pretraining on COVID-19 tweets. The BERTSCORE (DA) with BiLSTM model uses BERTSCORE (DA) to retrieve relevant misconceptions and a BiLSTM model for classifying tweet-misconception pairs. On the other side, BERTSCORE (DA) with SBERT (DA) model uses the combination of BERTSCORE (DA) and the Sentence-BERT representation with domain-adaptive pre-training for the classification of tweet-misconception pairs. Song et al. introduced a manually annotated COVID-19 disinformation corpus and proposed a model called CANTM for topic generation by taking into account the classification information regarding disinformation~\cite{Song2020}. They accumulated the properties of the BERT with a VAE model to build up a robust classification system. CANTM outperformed other baseline models in terms of accuracy and F1 score in the classification task. It also achieved the best perplexity score in the topic modeling task among all the models.

\subsection{Evaluation Metrics}
It is essential to compare the performance of the algorithms systematically. To evaluate the performance of algorithms, we used different metrics. These metrics include accuracy, precision, recall, F1 Score and many of the metrics have more than one name. All of these evaluation metrics are
derived from the values found in the confusion matrix which is a tabular representation of a classification model performance on the test set, which consists of four parameters: True Positive(TP), False Positive(FP), True Negative(TN), and False Negative(FN) based on the calculated
predicted class versus actual class(ground truth).

\subsubsection{Accuracy}
Accuracy is defined as the ratio of correctly predicted instances over the total number of evaluated instances. It is formally defined in Equation~\ref{eqn:acc}.
\begin{equation}
\label{eqn:acc}
    Accuracy = \frac{\text{TP}+\text{TN}}{\text{TP} + \text{FP}+ \text{TN}+ \text{FN}} 
\end{equation}

\subsubsection{Precision}
Precision is defined as the correctly predicted positive instances from the total predicted instances in a positive class. It is also known as positive predictive value (PPV). It is formally defined in Equation~\ref{equ:precision}.

\begin{equation}
\label{equ:precision}
    Precision = \frac{\text{TP}}{\text{TP} + \text{FP}} 
\end{equation}

\subsubsection{Recall}
Recall measures the fraction of positive instances that are correctly classified. It is also known as true positive rate (TPR) or sensitivity. It is formally defined in Equation~\ref{equ:recall}.

\begin{equation}
\label{equ:recall}
 Recall = \frac{\text{TP}}{\text{TP} + \text{FN}}
\end{equation}

\subsubsection{F1-score}
F1 Score is the weighted average of Precision and
Recall. It is formally defined in Equation~\ref{equ:f1-score}.

\begin{equation} 
\label{equ:f1-score}
F_{1} = 2\times{\frac{\text{Precision}\times{\text{Recall}}}{\text{Precision} + 
     \text{Recall}}}
\end{equation}

\subsection{Evaluation Results}
In the existing research on COVID-19 misinformation classification, several traditional ML and DL methods have been employed. Among them, some are highly efficient in the classification of COVID-19 misinformation and show higher  performance scores. Table~\ref{Best} represents the best performing models used in the existing studies on COVID-19 misinformation detection in terms of accuracy and F1-score.

\begin{table}[!ht]
  \centering
  \caption{Best performing models in terms of accuracy \& F1 score}
  \label{Best}
  \begin{adjustbox}{max width=\textwidth}
  \begin{tabular}{*{30}{c}}
 
  
 \hline
    Problem Tackled &Reference & Best Model&Train, Validation, Test ratio (\%) &Train size &Validation size&Test Size& \multicolumn{4}{c}{Performance Metrics}\\
    \cline{8-11}
    & & & & & & & Accuracy (\%)  & Precision (\%)  &Recall (\%) & F1 Score (\%)\\
    
  \hline
  \hline
  
  Rumor& \cite{Chen2020} & BERT & 10-fold C-V & - & -&- &99.20 &99.17 &98.13 &98.34\\
  \cline{2-11}
  & \cite{Alkhalifa2020}& CNN with CT-BERT & 97, 2, 1 & 9206 &150&140 &- &78 & - & -\\ 
   \cline{2-11}
   & ~\cite{Alsudias2020}&LR (COUNT VECTOR) &10-fold C-V &-&-&-&84.03 &81.04&80.03&80.5\\ 
   
   \hline
   
  Misleading Information &\cite{Elhadad2021f} & CNN &80,20,- & 5989 & 1497&- &99.99 &99.93 &100 &99.96\\ 
  \cline{2-11}
  
  &~\cite{Elhadad2020} &NN (TF) &5-fold C-V & - &-&-&99.68 &99.87&99.73&99.80\\ 
  \hline
  
  Fake News &\multirow{3}{*}{ \cite{Kar2020} }& mBERT\textunderscore NN(monolingual) &\multirow{3}{*}{80 , - , 20} & \multirow{3}{*}{1150} & \multirow{3}{*}{-}&\multirow{3}{*}{288} &-
   &87.17 &91.89 &89.47\\ 
   & &mBERT\textunderscore NN(multilingual)& & & & &- &76.47 &86.66 &81.25\\ 
  \cline{2-11}
  & \cite{Shahi2020}& BERT & - ,- ,- & - & - & - &- &78 &75 &76\\ 
 \cline{2-11}
   &\multirow{3}{*}{ \cite{Koirala2020} }& CNN &\multirow{3}{*}{54,26 ,20} & \multirow{3}{*}{1672} &\multirow{3}{*}{823}&\multirow{3}{*}{624} &80 &73 &70&72\\ 
   & &BERT& & & & &80&-& - & - \\ 
  \cline{2-11}
    &\cite{Dutta2020}  &  kNN (k-5) & - , -, - & - & -&- &89 &- &-&91 \\ 
   \cline{2-11}
   &\cite{Patwa2020}& SVM & 60 ,20 ,20& 6420&2140&2140 &93.46 &93.48 &93.46 &93.46\\
    \cline{2-11}
   &\cite{Zhou2020}& SAFE & 80 ,- ,20& - & - & -  &- & $75.15^{\bigstar}$ & $75.3 ^{\bigstar}$ & $75.25 ^{\bigstar}$\\ 
   \hline
 
  Misinformation & \cite{Cui2020} &dEFEND&75 , - , 25 &2,152 &- &717&-&89.65 &48.47&58.14\\
  \cline{2-11}
  
  &\multirow{3}{*}{ \cite{Carlos2020} }& RoBERTa(Classification of Users Comments) & 80 , - , 20 &2582 &-&645 &-&- &-
  & $90.3^{\bigstar}$ \\  
   & &SVM (Classification of YouTube Videos)&10-fold C-V &- & -&- &89.4 &-&-&-\\ 
  \cline{2-11}
  
      &\multirow{3}{*}{ \cite{Hossain} }&BERTSCORE (DA) + BiLSTM (SNLI)&\multirow{3}{*}{-, -, -} & \multirow{3}{*}{-} &\multirow{3}{*}{-}&\multirow{3}{*}{-} &-&44.2
   &45.3&43.1\\ 
   & &BERTSCORE (DA) + SBERT (DA) (MultiNLI)& & & & & -&55.9&50.9 &50.2\\ 
   & &BERTSCORE (DA) + SBERT (DA) (MedNLI)& & & & &-&47.8&49.2&48.4\\ 
     \cline{2-11}
 
   &\multirow{3}{*}{ \cite{Boukouvalas2020} }&$BERT_{BASE}$ (DNN)&\multirow{3}{*}{70, 30, -} & \multirow{3}{*}{392} &\multirow{3}{*}{168}&\multirow{3}{*}{-} &87.5 &84.7
   &90&87.3\\ 
   & &SVM/Gaussian (ICA)& & & & &81.2&85.9&76.3 &80.3\\ 
   & &SparseICA-EBM ($\lambda = 100$)& & & & &69.1&74.4&67.9&64.4\\ 
    \cline{2-11}
  
   &\multirow{3}{*}{ \cite{Al-Rakhami2020f} }& C4.5(Meta-model) &\multirow{3}{*}{10-fold C-V} & \multirow{3}{*}{-} &\multirow{3}{*}{-}&\multirow{3}{*}{-} &95.11 &95.3&95.1&95.1\\ 
   & & SVM+RF (Ensemble-model)& & & & &97.8&-&-&-\\
   \cline{2-11}
   
  & \cite{Kumar2020} &RoBERTa-large&- , - , - &- &- &- &-&73.75 &73.5&76.00\\ 
  \cline{2-11}
  &\cite{Perrio2020} &RoBERTa+PROB+TFIDF &70,10,20 &7000 &1000 &2000 &-&- &-&91.51\\ 
  \cline{2-11}
  
  &\multirow{3}{*}{ \cite{Dharawat2020} }&LR (Multiclass)&\multirow{3}{*}{80, -, 20} & \multirow{3}{*}{49,029} &\multirow{3}{*}{-}&\multirow{3}{*}{12,257} &96.3&31.3&23.3&25\\ 
   & &dEFEND w.news(Binary)& & & & &98 &92 &68 &75\\ 
    \hline
    
   Disinformation &\cite{Song2020}&CANTM&5-fold C-V&-&-&-&63.34&-& - & 55.48\\ 
   
   \cline{2-11}
    &\multirow{3}{*}{ \cite{Alam2020}}& BERT (En) (Binary) &\multirow{3}{*}{10-fold C-V }&\multirow{3}{*}{-}&\multirow{3}{*}{-}&\multirow{3}{*}{-}&- &- & - & $85.6^{\bigstar}$\\  
    
    & &mBERT (En) (Multiclass)& & & & &- &-&-& $53.48 ^{\bigstar}$\\ 
    & &mBERT (Ar) (Binary)& & & & &- &-&-& $83.96^{\bigstar}$\\ 
    & &FastText (Ar) (Multiclass)& & & & &- &-&-&$69.52^{\bigstar}$\\ 
    \hline
  &En=English & Ar=Arabic & $\bigstar=$ Macro~average  ~calculated &&&&&&&\\
\end{tabular}
\end{adjustbox}
\end{table}

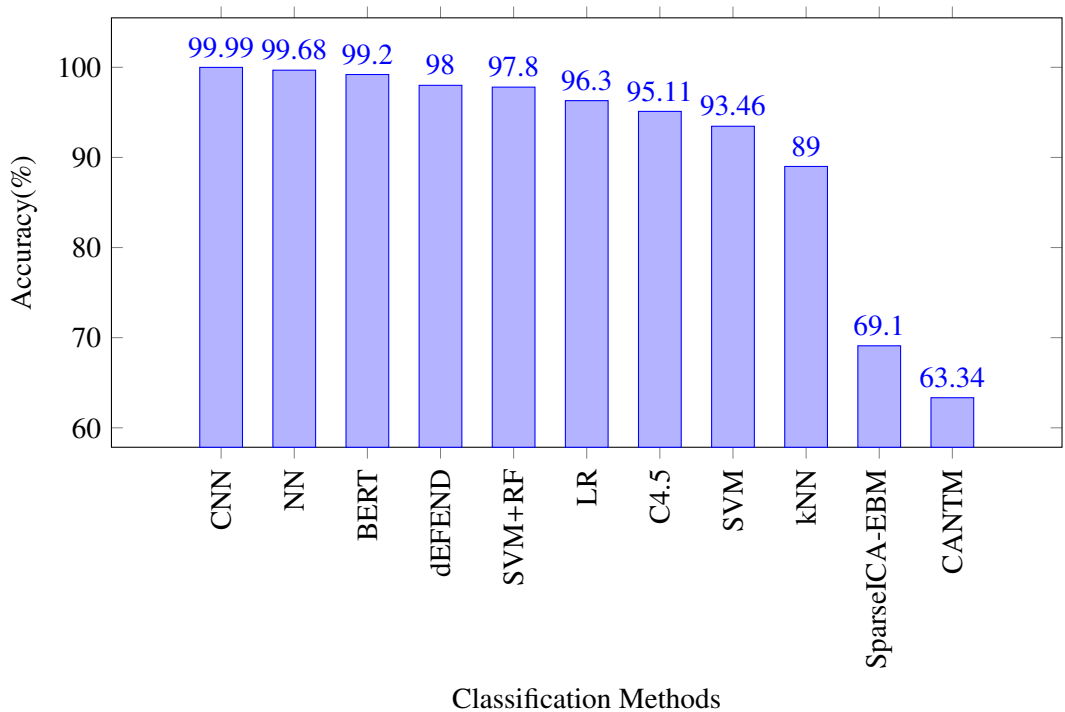
\begin{figure}[H]
\centering 
\begin{tikzpicture}
\begin{axis}[
ybar,
bar width=16pt,
x=2.5em,
enlargelimits=0.15,
legend style={at={(0.5,-0.2)},
anchor=north,legend columns=-1},
ylabel={Accuracy(\%)},
xlabel= {Classification Methods},
symbolic x coords={CNN,NN,BERT,dEFEND,SVM+RF,LR,C4.5,SVM,kNN,SparseICA-EBM,CANTM},
xtick=data, 
nodes near coords,
nodes near coords align={vertical},
x tick label style={rotate=90,anchor=east},
]
\addplot coordinates {
(CNN,99.99) (NN,99.68) (BERT,99.20) (dEFEND,98) (SVM+RF,97.8) (LR,96.3) (C4.5,95.11) (SVM,93.46)(kNN,89) (SparseICA-EBM,69.1)  (CANTM,63.34)
};
\end{axis}
\end{tikzpicture}
\caption{Best accuracy of various methods }
\label{Accuracy}
\end{figure}

We are able to see (from Table~\ref{Best}) that the CNN~\cite{Elhadad2021f} model obtained the highest score in terms of accuracy among all classification methods which is 99.99\% . This model used a train and test split of 80:20 without any validation split. The NN~\cite{Elhadad2020} model achieved an accuracy of 99.68\% using TF feature with 5-Fold C-V. BERT~\cite{Chen2020} with 10-fold C-V scored a 99.20\% accuracy among all the BERT models deployed in the existing studies. The dEFEND~\cite{Dharawat2020} model which uses co-attention on news content gave an accuracy of 98\%. This model also used a train and test ratio of 80:20 without any validation split. The ensemble model, SVM+RF \cite{Al-Rakhami2020f} obtained an accuracy of 97.8\% using 10-fold C-V. LR~\cite{Dharawat2020} showed 96.3\% accuracy which is the highest among all the models used in this study for a multi-class classification setting. Other methods such as C4.5~\cite{Al-Rakhami2020f}, SVM~\cite{Patwa2020}, kNN~\cite{Dutta2020}, SparseICA-EBM~\cite{Boukouvalas2020}, CANTM~\cite{Song2020} scored an accuracy of 95.11\%, 93.46\%, 89\%, 69.1\%,  63.34\% respectively. Figure~\ref{Accuracy} represents the best methods used in the existing studies in terms of accuracy.

\section{Open Issues \& Future Research Directions}
\label{sec:open issues}
During this COVID-19 pandemic, the propagation of misinformation through various platforms has already become a global concern. It has opened the door for the researchers to come up with different ideas to solve this problem. Accordingly, numerous researchers around the world are working on various research works on misinformation detection and classification related to COVID-19. In our systematic survey, we have presented the impact, characteristics, and detection of COVID-19 misinformation along with the research methodologies of the existing efforts. Researchers have proposed and implemented various techniques for the detection and classification of misinformation of COVID-19. Some of them are very efficient to classify misinformation with high accuracy value and some are not that much which can be taken into consideration for further improvements. Moreover, the number of notable works on COVID-19 misinformation detection is still not that big. Thus, we have pointed out several findings from these research works and the promising research directions for the future.

\subsection{Quality of the Datasets}
It is observed that there is still a lack of benchmark datasets that include resources to extract all relevant features related to COVID-19 misinformation. Besides, most of the studies have utilized the data that is mainly collected from social media platforms (e.g., Twitter, Facebook, etc.) and some other reliable sources. The majority of the datasets don't contain data from diverse sources. Moreover, class distribution in some data sets was observed to be imbalanced which affected the overall performance of the classification. Koirala showed that an increase in the weight of the minority class can handle this problem~\cite{Koirala2020}. A promising direction is to create a comprehensive, well-annotated, and large-scale benchmark dataset on COVID-19 misinformation which can be used by the scholars to conduct further research in this domain. Furthermore, future researchers may employ and investigate different sampling techniques to handle the class imbalance problem and demonstrate their effect on classification performance.

\subsection{Pre-processing \& Feature Selection on Large Volume Data }
In misinformation classification, data pre-processing is an underrated step. It was observed that most of the researchers give more focus on the method and often neglect the data pre-processing phase. Elhadad et al. showed that, with proper data-pre-processing approaches, the performance of the classification model can be significantly improved ~\cite{Elhadad2021f}. Usually, in the pre-processing step, special characters, punctuation marks, tags, URLs, stop words are removed and Part of Speech(PoS) tagging, word stemming, case-folding, etc are performed. In the future, researchers may work on dataset-specific pre-processing tasks. As the number of studies working with large volume COVID-19 misinformation data is relatively small, it was noticed that there are no efficient techniques for the selection of the important features on large-scale data. Future researchers may contribute to this scope by proposing methods on how to extract the most significant features from large volume data by minimizing the feature vector size effectively. 

\subsection{Multi-modality Based Detection System }
There is no study on COVID-19 misinformation detection that used multimodality such as texts, images, and videos altogether. Although individual modality is very important, it is not sufficient alone. Different modalities can help to gain different aspects of contents and derived information from different modalities complement each other to detect misinformation. The similarity between the image and the text is very important which can be an additional information for a comprehensive outcome. Thus, a study can be done by incorporating multimodal features to make a robust misinformation detection system. Though these multimodal systems can perform well in detecting misinformation, it can increase training and model size overhead, training cost \& complexity as the classifiers have always been trained with another classifier. In today’s competitive age, it is worthwhile to research those open issues and researchers can make contributions to solve these problems.

\subsection{Unsupervised Learning Based Techniques}
All the existing works on COVID-19 misinformation detection are supervised, which requires an extensive amount of time and a pre-annotated misinformation dataset to train a model. Obtaining a benchmark misinformation dataset on COVID-19 is also time consuming and labor-intensive work as the process needs careful checking of the contents. There is also a need to check other additional proof such as authoritative reports, fact checking websites, news reports etc. Leveraging a crowdsourcing approach to obtain annotations could relieve the burden of expert checking, but the  annotations’ quality may suffer~\cite{kim2018}. As misinformation is intentionally spread to mislead people, individual human workers alone may not have the domain expertise to differentiate real information and misinformation~\cite{Bond2006}. So it is time to consider semi-supervised or unsupervised models having limited or unlabeled data. Besides, unsupervised models can be more practical because unlabeled datasets are easier to obtain.

\subsection{Ensemble and Hybrid Learning Based Techniques}
A few works implemented ensemble methods to build more complex and effective models to better utilize extracted features. Ensemble methods build a conjunction of several weak classifiers to learn a stronger one that is more robust than any individual classifier alone. In the case of misinformation detection system, different variants of ensemble methods can significantly boost up the overall performance of the system. Again, hybrid classifiers (ML+ML, DL+DL) have been used for improving the predictions of the classification task in some existing literature~\cite{Al-Rakhami2020f,Song2020,Elhadad2021f,Hossain,Kumar2020}. Other combinations (ML+DL, DL+ML) of the hybrid classifier can be used for building up a robust classification system of COVID-19 misinformation.

\section{Conclusion}
\label{sec:conclusion}

In the course of COVID-19 pandemic, people are spending more time on internet to gather necessary information. Hence, when a piece of information is falsely represented, it spreads pretty fast and misguides its users by creating a strong negative impact on individuals and broader society. If we cannot halt the spread of COVID-19 misinformation, it may lead people to be more panicked by the overabundance of false information. To ease the detection of misinformation, traditional ML and DL methods are widely used to build up systems that can classify misinformation more precisely. In this survey, we outline various existing research works on COVID-19 misinformation classification and detection. In particular, we have provided a comprehensive view of different misinformation types and discussed existing methodologies to detect COVID-19 misinformation focusing on feature extraction methods, classification, detection performance etc. Comparing with the adopted existing techniques, DL appeared as one of the most efficient and effective techniques to classify misinformation accurately. Although sometimes the performance degrades, traditional ML methods also perform very well in the misinformation classification task. We also revealed the limitations of the existing studies and mentioned several research directions for further investigation in the future. We believe that our survey can provide important insights to build up robust classification systems for detecting misinformation related to COVID-19 and help researchers around the world to come up with new strategies to fight against the spread of misinformation during this pandemic.

\bibliographystyle{elsarticle-num-names} 
\bibliography{bibliography}

\begin{appendices}
\section{Abbreviation list}


\begin{tabular}{ll}
 ML& Machine Learning\\
 DL& Deep Learning\\
 NLP& Natural Language Processing\\
 C-V & Cross-Validation\\
 PCA& Principle Component Analysis\\
 ICA& Independent Component Analysis\\
 BoW & Bag of Words\\
TF-IDF & Term Frequency-Inverted Document Frequency\\
 LIWC&Linguistic Inquiry and Word Count\\
 RST& Rhetorical Structure Theory\\
 SVM&Support Vector Machine\\
 NB& Naive Bayes\\
 MNB& Multinomial Naive Bayes\\
 BNB& Bernoulli Naive Bayes\\
 kNN& k-Nearest Neighbors\\
 DT & Decision Tree\\
 RF & Random Forest \\
 ERF& Ensemble Random Forest\\
 LR& Logistic Regression\\
 GDBT& Gradient Boost\\
 BN& Bayes net\\
 MLP& Multi-Layer Perceptron\\
 NN& Neural Network\\
 CNN& Convolutional Neural Network \\
 RCNN& Recurrent Convolutional Neural Network\\
 RNN& Recurrent Neural Network \\
 LSTM& Long Short Term Memory \\
 BiLSTM & Bidirectional LSTM\\ 
 GRU& Gated Recurrent Unit \\
 BiGRU &Bidirectional GRU\\
  \end{tabular}
   
 \begin{tabular}{ll}
 BERT & Bidirectional Encoder Representations from Transformers\\
 CT-BERT & COVID-Twitter-BERT\\
 RoBERTa & Robustly optimized BERT approach\\
 ALBERT & A Lite BERT\\
 mBERT & multilingual BERT\\
 VAE& Variational Autoencoder \\
 SCHOLAR & Sparse Contextual Hidden and Observed Language Autoencoder \\
 HAN & Hierarchical Attention Networks\\
 dEFEND & Explainable FakE News Detection\\
 SAFE & Similarity-Aware FakE news detection \\
 SAME & Sentiment-Aware Multi-Modal Embedding \\
 CSI & Capture, Score, and Integrate \\
 CANTM&Classification Aware Neural Topic Model \\
\end{tabular}

\end{appendices}

\end{document}